\documentclass[aps,twocolumn,showpacs,floatfix,amssymb]{revtex4}
\usepackage{graphicx}
\begin{document}

\title{Spin excitations and thermodynamics of the\\
 $t$-$J$ model on the  honeycomb  lattice}

\author{ A.A. Vladimirov$^{a}$, D. Ihle$^{b}$ and  N. M. Plakida$^{a,c}$ }
 \affiliation{ $^a$Joint Institute for Nuclear Research,
141980 Dubna, Russia}
 \affiliation{$^{b}$ Institut f\"ur Theoretische Physik,
 Universit\"at Leipzig,  D-04109, Leipzig, Germany }
\affiliation{$^{c}$ Max-Planck-Institut f\"ur Physik Komplexer Systeme D-01187,
Dresden, Germany}
\thanks{\emph{ E-mail:} plakida@theor.jinr.ru
}%

\date{\today}

\begin{abstract}We present a spin-rotation-invariant Green-function theory for the dynamic spin
susceptibility in the spin-1/2 antiferromagnetic  $t$-$J$ Heisenberg model on the
honeycomb lattice. Employing a generalized mean-field approximation for arbitrary
temperatures and hole dopings, the electronic spectrum of excitations, the
spin-excitation spectrum and  thermodynamic quantities (two-spin correlation functions,
staggered magnetization, magnetic susceptibility, correlation length) are calculated by
solving a coupled system of self-consistency equations for the correlation functions. The
temperature and doping dependence of the magnetic (uniform static) susceptibility is
ascribed to antiferromagnetic short-range order. Our results on the doping dependencies
of the magnetization and susceptibility are analyzed in comparison with previous results
for the $t-J$ model on the square lattice.
\end{abstract}

\pacs{75.10.Jm ,75.10.-b, 75.40.Cx, 75.40.Gb}

\maketitle

\section{Introduction}
\label{sec:1}

 In recent years the two-dimensional carbon honeycomb lattice, the graphene,   has been
extensively  studied due its peculiar electronic properties (for a review
see~\cite{Neto09,Kotov12}). Studies of the graphene beyond the simple  model of
noninteracting electrons  by taking into account the Coulomb interaction  (CI) reveal a
rich phase diagram with phase transitions to the antiferromagnetic (AF) states,
spin-density wave (SDW), charge-density wave (CDW), and  nonconventional
superconductivity (SC).

In many papers the electronic properties of the Hubbard model  on the honeycomb lattice
were investigated. It was found that at a sufficiently  large  single-site Coulomb
repulsion  $U > U_c \approx 4 t $, the AF long-range order (LRO) emerges  for a single layer
close to half-filling ~\cite{Sorella92,Martelo97,Furukawa01}.  The phase diagram and spin
excitations of the Hubbard model for graphene  layers   using  the mean-field
approximation (MFA) and the random-phase approximation  were considered in
reference~\cite{Peres04}. Depending on the value of  $ U$ and electronic density $n$,
various phases were observed: at large $U > U_c \approx 3.8 t$, the  AF phase for $n
\lesssim 1$ was found, while at larger doping the ferromagnetic and  spiral phases were
obtained. The quantum phase transition in the half-filled Hubbard model on the honeycomb
lattice at $U > U_c$ with $U_c/t \approx 4-5 $ was found  in~\cite{Paiva05} using the
quantum Monte Carlo (QMC) and series expansion techniques. The temperature dependence of
the  specific heat also points to the AF phase transition at $U > U_c$. In
reference~\cite{Herbut06} phase transitions in  the Hubbard model of $N$-flavor electrons
on the  honeycomb lattice have been discussed in the limit of large $N$. There, a
semimetal to AF  insulator phase transition at the quantum critical point in the universality
class of the Gross-Neveu model was found. A general low-energy theory of electrons with
repulsive short-range CI on the honeycomb lattice at half-filling is presented
in~\cite{Herbut09}.

The phase diagram of extended Hubbard models with   nearest-neighbor (nn) and next
nearest-neighbor (nnn) repulsive interactions $V_1$  and $V_2$, respectively,  on the
honeycomb lattice  in  MFA was obtained in reference~\cite{Raghu08}. A phase transition
from the semimetal to Mott insulating phases at half-filling was found at large $U > U_c
\approx 3.8 t$. For small $V_1$ and $V_2$  the AF phase appears, while for larger $V_1$
and $V_2$ the renormalization group (RG) analysis  shows transitions to the SDW or CDW.

In a more recent QMC calculation ~\cite{Meng10} a gapped AF phase at half-filling for
$U/t > 4.3$  was  found, and for the intermediate coupling $3.5 < U / t < 4.3$, an
insulating gapped spin-liquid state formed by short-range resonating valence bonds  was
predicted.  But later QMC studies of larger clusters have not confirmed this transition
to the spin-liquid state~\cite{Sorella12}. The two-particle self-consistent approach  for
the Hubbard model on the honeycomb lattice in reference~\cite{Aria15} shows  the
semimetal to spin-liquid transition before the transition to the  AF state. In
reference~\cite{Yang12}, effective spin models for the Hubbard model on the honeycomb
lattice at half-filling were derived. It was observed that the six-spin interactions
frustrate the AF order and may lead the spin-liquid state behavior.  But the spin-liquid
state has not been found in other publications.  The transition from the  weak-coupling
semimetal to the strong-coupling insulating phase was studied in~\cite{Lang13} using  QMC
simulations for the SU$(N)$-symmetric Heisenberg model with the nn flavor exchange
interaction on the honeycomb lattice at half filling. In the SU(2) case a direct
transition between the semimetal and an AF insulator was obtained. In
reference~\cite{Assaad13} a continuous quantum phase transition between the semimetallic
and the insulating AF  states was found at $U_c/t =  3.78$  by considering a staggered
magnetic moment in the local magnetic field. A direct transition from a Dirac semimetal
to an AF Mott insulator was confirmed in reference~\cite{Toldin15} by using the
projective auxiliary-field QMC simulations and a finite-size scaling analysis. Although
the existence  of the spin-liquid state in the Hubbard model on the honeycomb lattice is
still under discussion, the transition from the semimetal to the AF LRO  phase  is proved
for a large enough  single-site CI, $U > U_c \approx 4 t$.

Superconducting phase transitions in the Hubbard model on the  honeycomb lattice have
been considered in several publications. The RG approach was used in~\cite{Honerkamp08}
to study phase transitions in the extended Hubbard model with the on-site interaction
$U$, the nn intersite repulsion $V$, and the spin-exchange interaction $J$. Close to
half-filling, the SDW or CDW orders occur for large $U$ and $V$, while  for a large
doping  $f$-wave triplet-pairing and $d+id$-wave singlet-pairing emerge. Chiral triplet
superconductivity on the graphene lattice was considered in~\cite{Faye15}.  Using the
dynamic cluster approximation for the Hubbard model with  $U/t = 2 - 6$,  a transition
from the $d + id$-wave singlet pairing at weak coupling  to the $p$-wave triplet pairing
at larger coupling was observed in~\cite{Xu16}. More references on studies of
superconducting phase transitions in the Hubbard model on the  honeycomb lattice may be
found in reference~\cite{Xu16}.

In the limit of strong correlations, $U \gg t$, the conduction band of electrons on the
honeycomb lattice  splits   into the singly- and doubly-occupied Hubbard subbands. In
this limit the Hubbard model can be reduced to the $t$-$J$ model for the projected
electron operators in one subband. This model was investigated by several authors.
In~\cite{Luscher06} a single-hole excitation  was considered within the $t$-$J$
spin-polaron model.  The results obtained for the honeycomb lattice are qualitatively
similar to those for the square lattice. A detailed study of the $t$-$J$ model on the
honeycomb lattice was presented in~\cite{Gu13}. The ground-state energy and the staggered
magnetization in the AF phase as function of doping $\delta$ have been calculated using
the Grassmann tensor product state approach, exact diagonalization and density-matrix
renormalization methods. The occurrence of the time-reversal symmetry breaking $d +
id$-wave SC at large doping was found. Moreover, a coexisting of the  SC and AF order was
observed for low doping, $ 0 <\delta< 0.1$, where the triplet pairing is induced (see
also~\cite{Gu14}).

In the  papers cited above mostly the phase diagram of the correlation  models on the
honeycomb lattice at zero temperature was studied. Less attention has been  paid to the
investigation  of electron- and spin-excitation spectra and of thermodynamic properties
as  functions  of temperature and electron concentration. Motivated by this shortcoming,
in the present paper we report results of  investigations of these spectra and of the
thermodynamics in the limit of strong correlations within the $t$-$J$ model. In our
previous paper~\cite{Vladimirov17} we have studied the honeycomb Heisenberg model at
half-filling over the whole temperature region both in the AF  and  paramagnetic phases.
Thereby, we have calculated the dynamic spin susceptibility (DSS) within the
spin-rotation-invariant  relaxation-function theory
~\cite{Vladimirov05,Vladimirov09,Vladimirov11} using the generalized mean-field
approximation (GMFA). Let us point out that the GMFA  has been successfully applied to
several quantum spin systems (see, e.g.,~\cite{Vladimirov17} and  references therein). In
the present paper we consider the effects of doping on AF order within the GMFA done for
the DSS. Similar studies have been done for the $t$-$J$ model on the square lattice in
our paper~\cite{Vladimirov09}.

In Section~\ref{sec:2} we formulate the $t$-$J$  model in terms of  Hubbard  operators.
The  electronic excitation spectrum is calculated in  Section~\ref{sec:3}.  The
spin-excitation spectrum and thermodynamic quantities are considered in
Section~\ref{sec:4}. The numerical results and discussion are given in
Section~\ref{sec:5}. The conclusion can be found in Section~\ref{sec:6}.

\section{The $t - J$ model }
\label{sec:2}

We study the  Hubbard model on the honeycomb lattice  in the limit of strong electron
correlations $U >> t$, when it  can  be reduced to the one-subband  $t$-$J$  model:
\begin{equation}
H =  - t\, \sum_{\langle i,j \rangle \sigma}  \tilde a_{i,\sigma}^{+} \tilde a_{j,\sigma}
- \mu \sum_{i , \sigma} \,n_{i,\sigma} + H_H, \label{b1}
\end{equation}
where $\tilde a_{i,\sigma}^{+} =a_{i,\sigma}^{+}(1-n_{i ,\bar{\sigma}})\;$ and $ \tilde
a_{i\sigma} =a_{i\sigma}(1-n_{i ,\bar{\sigma}})$ are projected creation and annihilation
electron operators with spin $\sigma/2$ ($\sigma=\pm 1 , \; \bar{\sigma} = - \sigma$) in
the singly occupied Hubbard subband, $n_{i ,\sigma}  = \tilde a_{i,\sigma}^{+}\, \tilde
a_{i,\sigma}$. Here, $t$ is the nn electron hopping energy.

The Heisenberg Hamiltonian  in  (\ref{b1}) is given by
\begin{equation}
 H_H  = \frac{J}{2}\sum_{\langle i, j \rangle}\;\left( {\bf S}_{i} \; {\bf S}_{j}
 - \frac{1}{4} n_i\, n_j \right),
 \label{b1a}
\end{equation}
where $J = 4t^2/U$ is the  nn  AF exchange interaction and $n_i = \sum_{ \sigma}
n_{i,\sigma}$.

To take into account on a rigorous basis the projected character of electron operators
$\tilde a_{i,\sigma}^{+} $, we employ the Hubbard operator (HO)
technique~\cite{Hubbard65}. The HOs  are defined as
\begin{equation}
X_{i}^{nm}=|i,n\rangle\langle i,m|  ,
 \label{b2}
\end{equation}
for three possible states at a lattice site $i$: $|i,n\rangle=|i,0\rangle$  and
$|i,\sigma\rangle$ for an empty site and for a singly occupied site by an electron with
spin $\sigma/2$, respectively.

The electron number operator and the spin operators in terms of HOs are defined as
\begin{eqnarray}
  n_i &=& \sum_{\sigma} X_{i}^{\sigma \sigma} =   X_{i}^{++}  +  X_{i}^{--},
\label{b3a}\\
S_{i}^{\sigma} & = & X_{i}^{\sigma\bar\sigma} ,\quad
 S_{i}^{z} =  (\sigma/2) \,[ X_{i}^{\sigma \sigma}  -
  X_{i}^{\bar\sigma \bar\sigma}] .
\label{b3b}
\end{eqnarray}
The completeness relation for the HOs, $\, X_{i}^{00} +
 \sum_{\sigma} X_{i}^{\sigma\sigma}  = 1 $,
rigorously preserves the constraint of no double occupancy of the quantum state
$|i,n\rangle $ on any lattice site $i$. From the multiplication rule $\, X_{i}^{nm}
X_{i}^{kl} = \delta_{mk} X_{i}^{nl} \,$ follow the commutation relations:
\begin{equation}
\left[X_{i}^{nm}, X_{j}^{kl}\right]_{\pm}= \delta_{ij}\left(\delta_{mk}X_{i}^{nl}\pm
\delta_{nl}X_{i}^{km}\right)\, .
 \label{b4}
\end{equation}
The upper sign  refers to  Fermi-type operators such as  $X_{i}^{0\sigma}$, while the
lower sign refers to  Bose-type operators such as  $n_i$  (\ref{b3a}) or the spin
operators (\ref{b3b}).
\par
Using the Hubbard operator representation, equation~(\ref{b2}) for $\tilde
a_{i\sigma}^{+} = X_{i}^{\sigma 0} \; $,  $\tilde a_{j\sigma}= X_{j}^{0\sigma}$ and
equations~(\ref{b3a}) and~(\ref{b3b}), we  write the Hamiltonian of the $t-J$ model
(\ref{b1}) in the form:
\begin{eqnarray}
H & = & - t\sum_{\langle i, j \rangle \sigma}\,  X_{i}^{\sigma 0}\, X_{j}^{0\sigma}
 - \mu \sum_{i \sigma} X_{i}^{\sigma \sigma}
\nonumber \\
&&
  +\frac{J}{4} \sum_{\langle i, j \rangle \sigma}\,
\left(X_i^{\sigma\bar{\sigma}}X_j^{\bar{\sigma}\sigma}  -
   X_i^{\sigma\sigma}X_j^{\bar{\sigma}\bar{\sigma}}\right).
\label{b5}
\end{eqnarray}
The Hamiltonian has the conventional form of the $t$-$J$ model in terms of Hubbard
operators (see, e.g.,~\cite{Plakida99}).

\begin{figure}
\centering
\resizebox{0.3\textwidth}{!}{\includegraphics{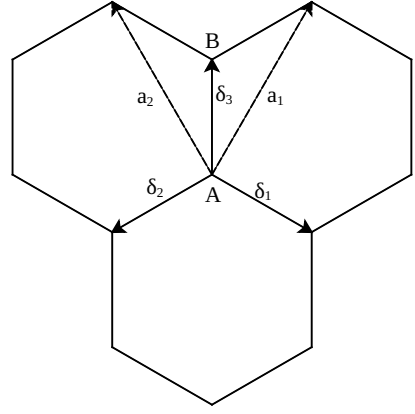}}
\caption{Sketch of the honeycomb lattice, where ${{\bf \delta}_1},\, {{\bf \delta}_2},\,
{{\bf \delta}_3} $ are the  nearest-neighbor vectors (\ref{nn}), and ${\bf a}_1$, ${\bf
a}_2$ are the lattice vectors. }
 \label{fig1}
\end{figure}

We consider the honeycomb lattice shown in Figure~\ref{fig1}. The lattice is bipartite
with two triangular sublattices $A$ and $B$. Each of the $N$ sites on the $A$ sublattice
is connected to three nn sites belonging to the $B$ sublattice by vectors ${{\bf
\delta}_j}$, and $N$ sites on $B$ are connected to $A$ by vectors $-{{\bf \delta}_j}$:
\begin{equation}
{{\bf \delta}_1} = \frac{a_0}{2}(\sqrt{3}, -1),\; {{\bf \delta}_2} = -
\frac{a_0}{2}(\sqrt{3}, 1), \; {{\bf \delta}_3} = a_0(0, 1). \label{nn}
\end{equation}
The basis vectors  are ${\bf a}_1 = {{\bf \delta}_3} - {{\bf \delta}_2} =
({a_0}/{2})(\sqrt{3}, 3)$ and ${\bf a}_2 = {{\bf \delta}_3} - {{\bf \delta}_1} =
({a_0}/{2})(-\sqrt{3}, 3)$, the lattice constant is $a = |{\bf a}_1| = |{\bf a}_2| =
\sqrt{3}a_0$, where $a_0$ is  the nn distance (see  Figure~\ref{fig1}); hereafter we put
$a_0 = 1$.  The  reciprocal lattice   vectors are ${\bf k}_1 = ({2\pi}/{3})(\sqrt{3}, 1)$
and $ {\bf k}_2 =({2\pi}/{3}) (-\sqrt{3}, 1)$. In the two-sublattice representation it is
convenient to split the site indices into the unit cell and sublattice indices, $i
\rightarrow  i \alpha$, $\alpha = A,\,B$.

The chemical potential $\mu$ depends on the average electron occupation number
\begin{equation}
  n = n_{\alpha} = \frac{1}{N} \sum_{ i, \sigma} \langle
 \, n_{i \alpha, \sigma}  \rangle ,
    \label{bn}
\end{equation}
where $N$ is the number of unit cells and $\langle ...\rangle$ denotes the statistical
average with the Hamiltonian (\ref{b5}).

\section{Electronic excitation spectrum}
\label{sec:3}

To calculate the electron excitation spectrum within the model (\ref{b5}), we consider
the anticommutator two-time  matrix Green function (GF)~\cite{Zubarev60}
\begin{eqnarray}
\hat{G}_{i j,\sigma} (t-t') & = &
  -i \theta(t-t') \langle \{ \hat{X}_i^{0 \sigma }(t) ,\hat{X}_j^{\sigma 0 }(t')\}\rangle
\nonumber\\
 & \equiv & \langle \!\langle \hat{X}_i^{0 \sigma }(t) ,\hat{X}_j^{\sigma 0 }(t')\rangle \!\rangle,
     \label{e1}
\end{eqnarray}
where $ \{ X, Y\} = XY + YX $,  $X(t)={\rm e}^{iHt} X {\rm e}^{-iHt}$, and $\,\theta(x)$
is the Heaviside function. Here we introduce the Hubbard operators in the two-sublattice
representation:
\begin{equation}
 \hat{X}_i^{0 \sigma }= \left( \begin{array}{c} X_{iA}^{0 \sigma } \\
X_{iB}^{ 0\sigma} \end{array} \right),\qquad \hat{X}_j^{\sigma 0}= \left(  X_{jA}^{\sigma
0 } X_{jB}^{\sigma 0}  \right) . \label{e2}
\end{equation}
The Fourier representation in $({\bf k}, \omega) $-space is defined by
\begin{eqnarray}
\hat{G}_{ij,\sigma} (t-t') &=& \int_{-\infty}^{\infty}\frac{d\omega }{2\pi}
e^{-i\omega(t-t')}
\nonumber\\
&\times&\frac{1}{N}\,
 \sum_{\bf k}{\rm e}^{i{\bf k ({\bf r}_i-{\bf r}_j)}} \hat{G}_{\sigma}({\bf k},\omega).
     \label{e1a}
\end{eqnarray}
Differentiating the GF (\ref{e1}) with respect to the time $t$ we get
\begin{equation}
\omega \, \hat {G}_{ij, \sigma} (\omega) = \delta_{ij} {\tau}_0 \, Q + \langle\!\langle
\hat{Z}_{i}^{0 \sigma } \mid \hat{X}_j^{\sigma 0} \rangle\!\rangle_{\omega}, \label{e3}
\end{equation}
where $\hat{Z}_{i}^{0 \sigma }= [ \hat{X}_i^{0 \sigma }, H]$, ${\tau}_0 $ is the unity
matrix, and $ Q = \langle   {X}_{i\alpha}^{ 0 0} + {X}_{i\alpha}^{ \sigma \sigma }
\rangle = 1-n_\alpha/2 $.

Now, we project the many--particle GF in (\ref{e3}) on the single--electron GF by
introducing the  irreducible  part of the  $\hat{Z}_{i}^{0 \sigma }$ operator,
\begin{eqnarray}
\langle\!\langle \hat{Z}_{i}^{0 \sigma } \mid  \hat{X}_j^{\sigma 0}
\rangle\!\rangle_\omega & = & \sum_l \hat{E}_{il\sigma} \langle\!\langle  \hat{X}_l^{0
\sigma } \mid \hat{X}_j^{\sigma 0} \rangle\!\rangle_\omega
\nonumber\\
 & + & \langle\!\langle \hat{Z}_{i}^{0 \sigma (irr)} \mid  \hat{X}_j^{\sigma 0}
\rangle\!\rangle_\omega  , \label{e4}
\end{eqnarray}
which  is orthogonal to the right-hand side operator: $\; \langle \{ \hat{Z}_{i}^{0
\sigma (irr)}, \hat{X}_j^{\sigma 0} \} \rangle = 0$. This results in the equation for the
frequency matrix,
\begin{equation}
\hat {E}_{ij} = \langle \{ [\hat{X}_i^{0 \sigma }, H], \hat{X}_j^{\sigma 0}\} \rangle \;
{Q}^{-1} \; . \label{e5}
\end{equation}
Using the Fourier transformation of the GF (\ref{e1a}) we obtain the equation for the GF
in the GMFA neglecting the last term in~(\ref{e4}) which describes inelastic scattering:
\begin{equation}
[\, \omega {\tau}_0 - \hat{E}({\bf k})] \, \hat {G}_\sigma ({\bf k}, \omega) = {\tau}_0\,
Q , \label{e6}
\end{equation}
where the electronic excitation spectrum  in GMFA is determined by the matrix of
correlation functions:
\begin{eqnarray}
  \hat{E}({\bf k})&=&
  \frac{1}{N}\sum_{i,j} \exp[i{\bf k (r_i-r_j)}] \hat{E}_{ij}
 \nonumber\\
   &= & {Q}^{-1} \left(
\begin{array}{cc}
\varepsilon ({\bf k})  &  \varepsilon_{AB}({\bf k}) \\
     \varepsilon_{AB}^*({\bf k}) & \varepsilon ({\bf k})
\end{array} \right) ,
\label{e7}
\end{eqnarray}
where $\varepsilon({\bf k}) = \langle \{ [ X_{{\bf k} A}^{0 \sigma }, H], X_{{\bf k}
A}^{\sigma 0}\} \rangle =
 \langle \{ [ X_{{\bf k} B}^{0 \sigma }, H],
X_{{\bf k} B}^{\sigma 0}\} \rangle$
 and $\varepsilon_{AB}({\bf k}) = \langle \{ [ X_{{\bf k} A}^{0 \sigma }, H],
X_{{\bf k} B}^{\sigma 0}\} \rangle$. The solution of the matrix equation for the GF
(\ref{e6})  reads:
\begin{eqnarray}
  \hat G_\sigma({\bf k}, \omega) = \left( \begin{array}{cc}
  G_{AA,\sigma}({\bf k}, \omega)\quad  G_{AB,\sigma}({\bf k}, \omega) \\
    G_{BA,\sigma}({\bf k}, \omega)\quad   G_{BB,\sigma}({\bf k}, \omega)\\
        \end{array}\right)
 \nonumber\\        =         \frac{Q}{D({\bf k}, \omega)} \left(
\begin{array}{cc}
 \varepsilon ({\bf k}) - \omega \quad - \varepsilon_{AB}({\bf k}) \\
   -  \varepsilon^*_{AB}({\bf k})\quad  \varepsilon ({\bf k}) - \omega\\
        \end{array}\right) .
\label{e8}
\end{eqnarray}
The electronic spectrum is defined from the equation
\begin{eqnarray}
D({\bf k}, \omega) & = &
  [\varepsilon ({\bf k}) - \omega]^2 - |\varepsilon_{AB}({\bf k})|^2
     \nonumber\\
 & = & [\varepsilon_{+}({\bf k}) - \omega] [\varepsilon_{-}({\bf k}) - \omega] ,
\end{eqnarray}
 and is given by
\begin{equation}
  \varepsilon_{\pm}({\bf k})  = \varepsilon ({\bf k}) \pm |\varepsilon_{AB}({\bf k})| .
      \label{e9}
\end{equation}

The calculation of the matrix elements in  (\ref{e7}) gives the following result for
$\varepsilon({\bf k})$:
\begin{eqnarray}
&&  \varepsilon({\bf k})  =
  - \mu +   \frac{3t}{Q}\, D_1 - \frac {3J}{4} n_\alpha +
\frac {3J}{2Q}\, C_1
 \equiv -\widetilde{ \mu }, \; \label{12a}
\end{eqnarray}
where we introduce the nn correlation functions   for electrons and spins,
\begin{eqnarray}
D_1 = \langle X^{\sigma 0}_{iA} X^{0\sigma}_{i+\delta_1, B}\rangle, \quad C_1 = \langle
S^{+}_{iA} {S}^{-}_{i+\delta_1, B} \rangle . \label{m7}
\end{eqnarray}
For the off-diagonal energy we have:
\begin{eqnarray}
&&\varepsilon_{AB}({\bf k}) = - \widetilde{t}\, \gamma_1 ({\bf k}),
 \label{12b}\\
    & & \widetilde{t}  =  t\, Q \, [1 + \frac{3 C_1}{2 Q^2}]
   + J\frac { D_1}{2 Q},
\label{12c}
\end{eqnarray}
where $\gamma_1({\bf k}) = \sum_b \exp (i {\bf k} \overrightarrow{\delta_b})$ and $
|\gamma_1({\bf k})|^2 = {1 + 4 \cos ( \sqrt{3}k_x /2)[\cos (\sqrt{3} k_x /2) + \cos
({3}k_y/{2})]} $. Note that equations~(\ref{12a}), (\ref{12b}), (\ref{12c}) are similar
to the results obtained for the spectrum of the  $t$-$J$ model on the square lattice
in~\cite{Plakida99}.

Thus, the electronic spectrum has two branches:
\begin{eqnarray}
\varepsilon_{\pm}({\bf k}) = -\widetilde{\mu} \pm  \widetilde{t}\,|\gamma_1({\bf k})|.
\label{12d}
\end{eqnarray}
It agrees with the spectrum of graphene (see, e.g.,~\cite{Neto09,Wallace47}), except for
the renormalization of the chemical potential  $\widetilde{ \mu }$ and the hopping
parameter  $\widetilde{t}$ due to strong correlations. Therefore, in the
strong-correlation limit the cone-type dispersion is conserved, i.e., the spectrum
reveals Dirac cones at the corners (K points) of the Brillouin zone (BZ).

\begin{figure}
\centering
\resizebox{0.3\textwidth}{!}{%
\includegraphics{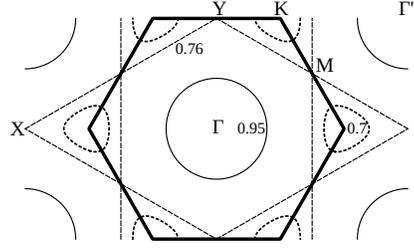}}
\caption{Brillouin zone (bold) and hole  Fermi surface for  $n = 0.95$
(thin solid), $0.76$ (dashed), and $0.7$ (dotted ).}
 \label{fig2}
\end{figure}

The BZ and the Fermi surfaces (FS) for holes  at the electronic occupation numbers $n =
0.95, \; 0.76,\;$ and $0.7$ are shown in Figure~\ref{fig2}. At $n \lesssim 1$ the hole FS
is small and centered at the $\Gamma$ point. With decreasing $n$ the FS becomes larger,
and at some characteristic value  $n_0 = 0.76 $ the  FS touches the BZ at M-points. At
larger hole doping,  six  pockets centered at the $K$-points emerge which shrink to
points for the half-filled band at $n = 2/3$.

For the diagonal  GF we have
\begin{eqnarray}
G_{\alpha \alpha, \sigma}({\bf k},\omega)
 = \frac{Q}{2 }\,\left[ \frac{ 1}
 {\varepsilon_{+}({\bf k}) - \omega}  + \frac{1}
{\varepsilon_{-}({\bf k}) - \omega}\right] . \label{e10}
\end{eqnarray}
The mean occupation number of electrons is equal to
\begin{eqnarray}
 n_{\alpha} = \frac{1}{N}\sum_{{\bf k}, \sigma} n_{\alpha \sigma}({\bf k}),
 \label{e11b}
\end{eqnarray}
with
\begin{eqnarray}
&&  n_{\alpha \sigma}({\bf k}) =\langle X_{{\bf k}\alpha }^{\sigma \sigma} \rangle =
\langle X_{{\bf k}\alpha }^{\sigma 0}  X_{{\bf k}\alpha }^{0 \sigma} \rangle
\nonumber\\
 & & =\int^{\infty}_{-\infty} \frac{d\omega}{e^{\omega/T} +1}\, [-\frac{1}{\pi} \mbox{Im}
G_{\alpha \alpha, \sigma}({\bf k},\omega)] \nonumber\\
& & = Q \,\frac{1}{2 }[N(\varepsilon_{+}({\bf k}))+ N(\varepsilon_{-}({\bf k}))],
 \label{e11}
\end{eqnarray}
where $  N(\varepsilon_{\pm}({\bf k})) =  [\exp[\varepsilon_{\pm}({\bf k})/T]+1]^{-1}  $,
and $ n_{\alpha}  \leq 1$.

For the off-diagonal GF in~(\ref{e8}) we obtain
\begin{eqnarray}
&& G_{AB, \sigma}({\bf k},\omega) = Q \frac{\varepsilon_{AB}({\bf k})}{2\,
|\varepsilon_{AB}({\bf k})|}
\label{e10b}\\
 & &
\times \Big[\frac{1}{\varepsilon_{+}({\bf k}) - \omega} - \frac{1}{\varepsilon_{-}({\bf
k}) - \omega}\Big],
 \nonumber
\end{eqnarray}
and for the corresponding correlation function we get
\begin{eqnarray}
&&  \langle X_{{\bf k} B }^{\sigma 0}  X_{{\bf k} A}^{0 \sigma} \rangle
 =  Q\,\frac{\gamma_1({\bf k})}
   {2|\gamma_1({\bf k})|}
 \nonumber\\
    & & \times [  N(\varepsilon_{-}({\bf k}))  - N(\varepsilon_{+}({\bf k}))] .
 \label{e12}
\end{eqnarray}

\section{Spin-excitation spectrum and thermodynamics}

\label{sec:4}

To calculate the spin-excitation spectrum  and to evaluate the thermodynamic quantities
in the model  (\ref{b5}), we consider the  two-time matrix commutator
GF~\cite{Zubarev60}:
\begin{eqnarray}
 \hat{G}^\pm_{ij} (t-t')& = &  -i \theta(t-t') \langle [{S}^+ _{i}(t) \; , \; {S}^- _{j}(t') ]\rangle
\nonumber\\
 & \equiv & \langle \!\langle {S}^+_{i}(t) \mid {S}^-_{j }(t')\rangle \!\rangle,
     \label{g2}
\end{eqnarray}
where $ [S, Y] = SY - YS$.  The Fourier representation of the spin GF, $\langle \!\langle
{S}^+_{{\bf q}} \mid {S}^-_{-{\bf q}}\rangle \!\rangle_\omega $, is defined by the same
relation as for the electronic GF (\ref{e1a}).

In the relaxation-function theory developed on the basis of the equation of motion method
in~\cite{Vladimirov09,Vladimirov05} we obtain the following representation of the DSS
$\hat\chi({\bf q},\omega) = -\hat{G}^{+-}({\bf q},\omega)$:
\begin{equation}
\hat\chi ({\bf q},\omega) = [ \hat F({\bf q}) + \hat \Sigma({\bf q},\omega)  - \omega^2
\, \tau_0 \, ]^{-1}\times \hat m ({\bf q}).
     \label{g4}
\end{equation}
Here, $\hat F({\bf q})$  is the  frequency matrix of spin excitations in the GMFA, where
the approximation $-\ddot{S}^{+}_{{\bf q}\alpha}= [[S^{+}_{\bf q \alpha}, H],
H]=\Sigma_{\beta} F_{\alpha \beta}({\bf q}) S^{+}_{{\bf q}\beta}$ is made, and $\hat m
({\bf q})$ is the moment matrix with components
 $m_{\alpha \beta}({\bf q}) = \langle [i \dot{S}^{+}_{{\bf q}\alpha} ,
S^{-}_{-{\bf q}\beta}] \rangle = \langle [[S^{+}_{{\bf q}\alpha}, H], S^{-}_{-{\bf
q}\beta}] \rangle$. The self-energy $\hat \Sigma({\bf q},\omega)$ can be expressed
exactly by a multispin GF (see~\cite{Vladimirov09,Vladimirov05}).

We consider the GMFA  for the DSS neglecting the self-energy in~(\ref {g4}). Then, for a
lattice with basis the zero-order DSS is given by (cf.~\cite{Vladimirov17})
\begin{equation}
[\hat F({\bf q})  - \omega^2 \, \tau_0 \, ] \times
 \hat \chi ({\bf q},\omega) = \hat m({\bf q}).
     \label{g6}
\end{equation}
For the static spin susceptibility we obtain
 \begin{equation}
\hat \chi ({\bf q}, 0) \equiv \hat \chi ({\bf q}) = \hat F^{-1}({\bf q}) \times \hat
m({\bf q}).
 \label{g6a}
\end{equation}
The direct calculation  of the matrix elements $m_{\alpha \beta}({\bf q})$ yields
\begin{equation}
\hat m ({\bf q}) = \left(  \begin{array}{cc} m_{AA} & m_{AB}({\bf q})\\
  m_{AB}^*({\bf q}) & m_{AA} \end{array}  \right),
 \label{b7}
 \end{equation}
where
\begin{eqnarray}
&&m_{AA} = -6 J C_1 + 6 t D_1,
 \label{b7a}\\
&&m_{AB}({\bf q}) = (2 J C_1 - 2 t D_1) \gamma_1({\bf q}) . \label{b7b}
\end{eqnarray}

To calculate  the frequency matrix $\hat{F}({\bf q})$ in equation~(\ref {g6}), we start
from the second derivative $-\ddot{S}^{+}_{i}$. Taking into account only the diagonal
contributions $\, F_{i}^{tt} = [\, [ {S}^{+}_{i} , H_t] , H_t]\,$, and $\, F_{i}^{JJ} =
[\, [ {S}^{+}_{i} ,H_J] , H_J ]\,$,
 where $H_t (H_J)$ is the hopping (exchange) part of the model (\ref {b5}), we obtain
\begin{eqnarray}
F_{i}^{tt}&= &\sum_{j,n}t_{ij}\Bigl\{t_{jn}\left[H^-_{ijn}+H^+_{nji}\right]
-  (i \Longleftrightarrow j ) \Bigr\},  \label{A2} \\
F_{i}^{JJ} & = & \frac{1}{4}\sum_{j,n}J_{ij}\Bigl\{J_{jn}\left[2P_{ijn}+\Pi_{ijn}\right]
  - (i \Longleftrightarrow j )\Bigr\},  \label{A3}
\end{eqnarray}
where
\begin{eqnarray}
H^{\sigma}_{ijn}&=&X_{i}^{\sigma 0}X_{j}^{+-}X_{n}^{0\sigma} +
X_{i}^{+0}(X_{j}^{00}+X_{j}^{\sigma\sigma})X_{n}^{0-},
 \label{A4}\\
P_{ijn}&=&S_{i}^{z}S_{j}^{z}S_{n}^{+} + S_{n}^{+}S_{i}^{z}S_{j}^{z} \nonumber\\
 & - & S_{i}^{z}S_{j}^{+}S_{n}^{z} - S_{n}^{z}S_{i}^{z}S_{j}^{+},
 \label{A5}\\
\Pi_{ijn}&=&S_{i}^{+}S_{j}^{-}S_{n}^{+}+S_{n}^{+}S_{i}^{+}S_{j}^{-}
\nonumber\\
& -& S_{i}^{+}S_{j}^{+}S_{n}^{-}-S_{n}^{-}S_{i}^{+}S_{j}^{+} \, .
 \label{A6}
\end{eqnarray}
We do not consider the off-diagonal terms $F_{i}^{tJ}, F_{i}^{Jt}$ as discussed
in~\cite{Vladimirov09}.

We perform the following  decoupling procedure preserving the  local correlations.
Decoupling the operators  in $H^{\sigma}_{i j n}$ we introduce the parameter $\lambda$:
\begin{equation}
X_{i}^{\sigma 0}X_{j}^{+-}X_{n}^{0 \sigma} =
   \lambda \langle\, X_{i}^{\sigma 0} X_{n}^{0 \sigma}\rangle \, S_{j}^{+},
 \label{A7}
\end{equation}
where for $\, n = i,\, $  $ \,\langle X_{i}^{\sigma 0} X_{i}^{0 \sigma}\rangle =\langle
X_{i}^{\sigma \sigma}\rangle = n/2\,$, and the second term of $\,H^{\sigma}_{ijn}\,$ with
$ n \neq  i$  is neglected (cf.~\cite{Winterfeldt98}). Analogously, the  operators in
$\Pi_{i j n= i}$ and $ P_{ij, n= i}$ are decoupled as
\begin{equation}
\Pi_{i j i} = 2P_{i j i}  = - (1 - X_{i}^{00})\,S_{j}^{+} = - (1 - \lambda\,\delta)
S_{j}^{+},
 \label{A8}
\end{equation}
where $\delta = \langle X_i^{00} \rangle = 1 - n$ is the hole concentration, and we used
the equations: $ \, S_{i}^{+}S_{i}^{-} = X_{i}^{++},\; S_{i}^{z}S_{i}^{+} + S_{i}^{+}
S_{i}^{z}  = 0, \,$ and $\, ( S_{i}^{z})^2 = (1/4)(X_{i}^{++} + X_{i}^{--})\,  $. The
parameter $\lambda$ describes the renormalization  of the vertex for spin scattering on
charge fluctuations.

The conribution $F_i^{JJ} $  is proportional to  products of three spin operators  on
different lattice sites along nn sequences, e.g., $\langle i A, j B, k A \rangle $. We
perform the decoupling of them as follows
\begin{eqnarray}
S^+_{iA} S^z_{jB} S^z_{kA} &  = & \alpha_1 \langle S^z_{jB} S^z_{kA} \rangle S^+_{iA} =
\frac{\alpha_1}{2} C_{1}\,S^+_{iA},
\label{b12a}\\
S^+_{jB}S^z_{iA} S^z_{kA} & = & \alpha_2 \langle S^z_{iA} S^z_{kA} \rangle S^+_{jB}
 = \frac{\alpha_2}{2} C_{2} \, S^+_{jB},
 \label{b12b}\\
S^+_{iA} S^+_{jB} S^-_{kA} & =  & \alpha_1 \langle S^+_{jB} S^-_{kA} \rangle S^+_{iA} + \alpha_2 \langle S^+_{iA} S^-_{kA} \rangle S^+_{jB} \nonumber\\
& = & \alpha_1 C_{1}\,S^+_{iA} + \alpha_2 C_{2} \, S^+_{jB}. \label{b12c}
\end{eqnarray}
 Here, the vertex renormalization parameters $\alpha_1$ and $\alpha_2$ are
attached to the nn and the nnn  correlation functions $C_1$ and $C_2$, respectively, and
the equality $\langle S^z_i S^z_j \rangle = \frac{1}{2} \langle S^+_i S^-_j \rangle$ due
to spin-rotation invariance is taken into account.
 Using these decouplings we obtain the frequency matrix $\hat F({\bf q})$:
\begin{equation}
\hat F({\bf q}) = \frac{1}{2}\left(  \begin{array}{cc} F_{AA}({\bf q}) & F_{AB}({\bf q})
\\ F^*_{AB}({\bf q}) & F_{AA}({\bf q}) \end{array} \right) ,
\label{b13}
\end{equation}
where
\begin{eqnarray}
F_{AA}({\bf q}) & = &  J^2 (3(1 - \lambda\delta) + 12 \alpha_2 C_2 + 2 \gamma_2({\bf q})
\alpha_1 C_1)
  \nonumber \\
 & + & 3 t^2 \lambda (\delta - 2D_2),
  \label{b14} \\
F_{AB}({\bf q}) & = & -  \gamma_1({\bf q})\, f_{AB},   \label{b14a}\\
f_{AB} & = &
 J^2 (1 - \lambda\delta + 4 \alpha_1 C_1 + 4 \alpha_2 C_2) \nonumber\\
 & + & t^2 \lambda (\delta - 2D_2),
  \label{b14}
\end{eqnarray}
with
\begin{eqnarray}
\gamma_2({\bf q}) &=& \sum_{i, j\neq i} \exp (i {\bf q}
({{\bf \delta}_i}-{{\bf \delta}_j})) \nonumber\\
&=& 4 \cos (\frac{\sqrt{3}}{2} q_x) \cos (\frac{3}{2} q_y) + 2 \cos (\sqrt{3} q_x),
\label{b15}
\end{eqnarray}
and the nnn correlation function $D_2 = \langle X^{\sigma 0}_{i \alpha}
X^{0\sigma}_{i+{\bf a}_1, \alpha} \rangle$.

Since the  matrices $\hat m({\bf q})$ and $\hat F({\bf q})$ commute, it is convenient to
solve equation~(\ref {g6}) by introducing the eigenvalues  $m_{\pm}({\bf q})$ and the
normalized eigenvectors $|E_{\pm}({\bf q}) \rangle$ of the matrix~(\ref {b7}):
\begin{equation}
[\hat m({\bf q}) - \hat \tau_0 m_{\pm}({\bf q})] |E_{\pm}({\bf q})\rangle = 0 ,
\label{b10}
\end{equation}
which are given by
\begin{eqnarray}
m_{\pm}({\bf q})& = & - (2 J C_1 - 2 t D_1) (3 \pm |\gamma_1(q)|)  ,
\label{b11}\\
|E_{\pm}({\bf q})\rangle & =& \frac{1}{\sqrt{2}}\left(  \begin{array}{cc} \mp \gamma_1
({\bf q}) / |\gamma_1 ({\bf q})|\\ 1 \end{array}  \right). \label{b11a}
\end{eqnarray}
For the same eigenvectors  $|E_{\pm}({\bf q})\rangle$  the spin-excitation frequencies
are obtained as
\begin{equation}
\omega^2_{\pm}({\bf q}) = \frac{1}{2}(F_{AA}({\bf q}) \pm |  \gamma_1({\bf q}) |f_{AB}).
 \label{b17}
\end{equation}
In this notation the DSS reads
\begin{equation}
\chi_{\alpha \beta} ({\bf q},\omega) = \sum_{j = \pm} \chi_{j} ({\bf q}, \omega)
 \langle \alpha | E_j({\bf q}) \rangle \langle
E_j({\bf q}) | \beta \rangle,
 \label{b16}
\end{equation}
where
\begin{equation}
\chi_{\pm} ({\bf q}, \omega) = \frac{m_{\pm}( {\bf q})} {\omega_{\pm}^2({\bf q}) -
\omega^2},
\end{equation}
and $\langle \alpha | E_{\pm} \rangle\langle E_{\pm} | \beta\rangle = 1/2$ for $\alpha =
\beta$,  otherwise $\langle \alpha | E_{\pm} \rangle\langle E_{\pm} | \beta\rangle = \mp
\gamma_1({\bf q}) / (2|\gamma_1({\bf q})|)$.

Using the spectral representation for the GF the correlation function $ C_{{\bf r} \alpha
\beta} = \langle S^+_{0 \alpha} {S}^-_{{\bf r} \beta} \rangle$ is written as
\begin{eqnarray}
 C_{{\bf r} \alpha \beta} =
\frac{1}{N}\, \sum_{{\bf q}  \neq {\bf Q}}  C_{\alpha\beta}({\bf q}) {\rm e}^{i{\bf q r}}
+ C_{\alpha \beta} {\rm e}^{i {\bf Q r}},
     \label{b18}
\end{eqnarray}
where
\begin{eqnarray}
C_{\alpha \beta}({\bf q}) &=& \sum_{j = \pm} \frac{m_{j} ({\bf q})}{2 \omega_{j}({\bf
q})} \coth
\frac{\omega_{j}({\bf q})} { 2 T} \nonumber\\
&\times & \langle \alpha | E_j({\bf q}) \rangle \langle E_j({\bf q})| \beta \rangle,
 \label{b19}
\end{eqnarray}
$C_{\alpha\alpha} = - C_{\alpha \neq \beta} =  C$, and  the wave  vector ${\bf Q}$
characterizes the long-range order (LRO).  The condensation part $C$ appears in the
ordered phase when $\omega_+({\bf q})$ condensates at ${\bf Q}$  which determines the
LRO, $\omega_+({\bf Q}) = 0$. In the  case of AF  order in the two-sublattice model, we
have
 ${\bf Q} = (0, 0)$, and the staggered magnetization $m_{st}$ is determined  by
\begin{equation}
(m_{st})^2 = \frac{3}{2} \, C.
 \label{b17a}
\end{equation}
Let us consider the uniform static susceptibility $\chi = \frac{1}{2}(\chi_{AA}(0) +
\chi_{AB}(0)) = \frac{1}{2}\chi_-(0)$ and the staggered susceptibility $\chi_{st} =
\chi_{AA}({\bf Q}) - \chi_{AB}({\bf Q}) = \chi_+({\bf Q})$. Expanding $\chi_-({\bf q})$
around ${\bf q} = 0$ we obtain
\begin{equation}
 \chi = \frac{- 2 J C_1 + 2 t D_1}{J^2 (1 - \lambda\delta + \alpha_1 \,C_1 +
 4 \alpha_2 \,C_2) + t^2 \lambda (\delta - 2 D_2)}.
 \label{b19a}
\end{equation}

We expand $\chi_+({\bf q})$ in the neighborhood of the AF vector ${\bf Q}$ and obtain
$\chi_+({\bf Q} + {\bf k}) = \chi_+({\bf Q})[1+\xi^2 (k_x^2 + k_y^2) ]^{-1}$, where for
the  correlation length $\xi$ we get:
\begin{equation}
\xi^2 = - \frac{3[J^2(1 - \lambda\delta + 7  \alpha_1 C_1 + 4  \alpha_2 C_{2}) + t^2
\lambda (\delta - 2 D_2)]}{2\omega_+^2({\bf Q})}  .
 \label{b20a}
\end{equation}
At zero temperature, for $\delta < \delta_c$ the LRO occurs when both the correlation
length (\ref{b20a}) and $\chi_+({\bf Q})$ diverge.

\section{Results}
\label{sec:5}

To evaluate the spin-excitation spectrum and the thermodynamic properties, the
correlation functions $C_1$, $C_2$, the transfer amplitudes $D_1$, $D_2$, and the vertex
parameters $\alpha_1, \, \alpha_2,\,$ and $ \lambda, \, $ appearing in the spectrum
$\omega_{\pm}({\bf q})$ as well as the condensation term $C$ in the LRO phase have to be
determined as solution of a coupled system of self-consistency equations. Besides
equations~(\ref{b18}) and (\ref{e12}) for calculating the correlation functions and the
transfer amplitudes, respectively, we have the sum rule $C_{{\bf r} = 0, \alpha \alpha} =
\langle {S}^{+}_{i\alpha} {S}^{-}_{i\alpha} \rangle = (1-\delta)/2$ and the LRO condition
$\omega_{+}({\bf Q}) = 0$. That is, we have more parameters than equations. To obtain a
closed system of equations, we take the  following choice of the vertex parameters.

\begin{figure}
\centering \resizebox{0.4\textwidth}{!}{\includegraphics{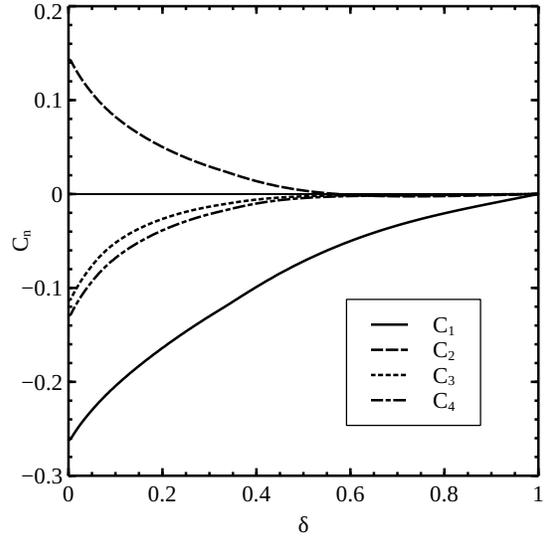}}
 \caption[]{Spin correlation functions $C_n$ between the n-th (n = 1,2,3,4)
nearest neighbors vs doping at $T = 0$ and $J/t = 1/3$. }
 \label{fig3}
\end{figure}

\begin{figure}
\centering \resizebox{0.4\textwidth}{!}{\includegraphics{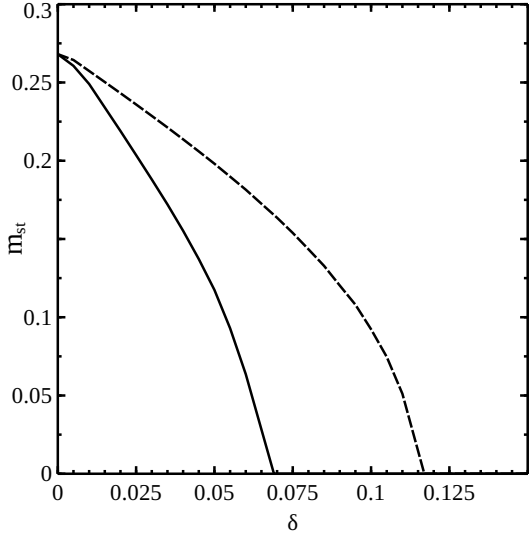}}
 \caption[]{Staggered magnetization $m_{st}$ at $T = 0$ as a function
of doping for $J/t = 1/3$ (solid) and $J/t = 1/2$ (dashed).}
 \label{fig4}
\end{figure}

\begin{figure}
\centering \resizebox{0.4\textwidth}{!}{\includegraphics{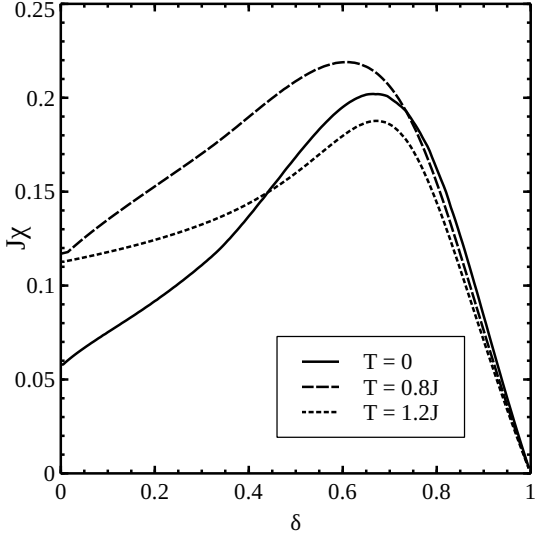}}
 \caption[]{Uniform static spin susceptibility $J\chi$ vs
doping at $J/t = 1/3$.}
 \label{fig5}
\end{figure}

\begin{figure}
\centering \resizebox{0.4\textwidth}{!}{\includegraphics{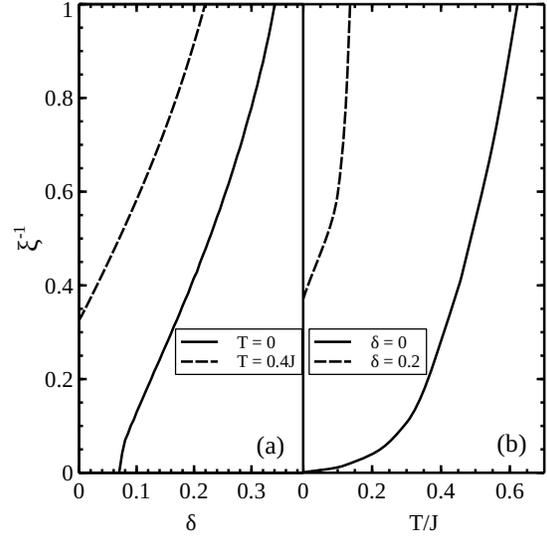}}
 \caption[]{Inverse correlation length $\xi^{-1}$ at $J/t = 1/3$
as a function of doping (a) and temperature (b).}
 \label{fig6}
\end{figure}

\begin{figure}
\centering \resizebox{0.4\textwidth}{!}{\includegraphics{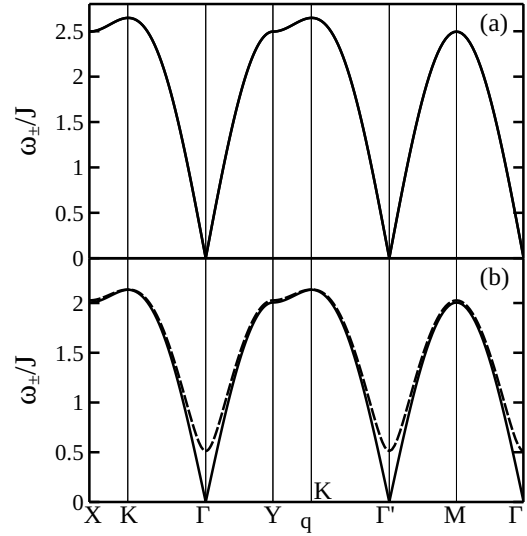}}
\caption[]{Spin-excitation spectrum  $\omega_-({\bf q})$ (solid) and $\omega_+({\bf q})$
(dashed) at $T = 0$ for (a) $\delta = 0$ and (b) $\delta = 0.2$.}
 \label{fig7}
\end{figure}

First, let us mention that the parameters $\alpha_1$ and $\alpha_2$ cannot be used to
satisfy the sum rule at high enough hole densities, because for $t \gtrsim J$ the $t^2$
term in~(\ref{b14}) will dominate the $J^2$ term at high enough doping levels. Therefore,
the influence of $\alpha_{1,2}$ on $C_{0,\alpha\alpha}$ rapidly weakens with increasing
doping. The vertex parameter $\lambda$, however, which describes the coupling between the
spin and hole degrees of freedom, is suitable to satisfy the sum rule over the whole
doping region. Therefore, we determine the parameters $\alpha_1$ and $\alpha_2$ in the
Heisenberg limit $(\delta = 0)$ and take their values also at finite $\delta$, as was
also done in~\cite{Vladimirov09,Vladimirov11}. Following reference~\cite{Vladimirov17} we
fix $\alpha_1$ and $\alpha_2$ by the sum rule $C_{0,\alpha\alpha} = 1/2$ and by the QMC
value of the staggered magnetization, $m_{st}(0) = 0.2681$, given in~\cite{Low09}. We get
$\alpha_1(0) = 2.91$ and $\alpha_2(0) = 3.57$. At finite temperatures we determine
$\alpha_1(T)$ and $\alpha_2(T)$ from the sum rule and the ansatz $r_{\alpha}(T) \equiv
[\alpha_2(T) - 1]/[\alpha_1(T) - 1] = r_{\alpha}(0)$
(see~\cite{Vladimirov17,ST91,WI97,Junger09}). At finite doping, we use the sum rule
$C_{0,\alpha\alpha} = (1-\delta)/2$ to calculate $\lambda(T,\delta)$.

In Figure~\ref{fig3} our results for the doping dependence of the spin correlation
functions $C_n$ up to $n = 4$ at $T = 0$ are presented. The different sign of $C_n$
reflects the AF order which gradually decreases with increasing doping, where at $\delta
\gtrsim 0.6$ only nn spin correlations survive.

Considering the staggered magnetization $m_{st}(\delta)$ at $T = 0$ depicted in
Figure~\ref{fig4}, the AF LRO is suppressed with increasing doping due to the spin-hole
interaction. At the critical doping $\delta_c(J/t)$ we obtain a smooth phase transition
from the LRO phase to a paramagnetic phase with AF SRO, where $\delta_c(J/t = 1/3) =
0.069$ and $\delta_c(J/t = 1/2) = 0.117$. For $J/t = 1/3$ the critical doping lies near
the value $\delta_c \gtrsim 0.1$ found in~\cite{Gu13} by density-matrix renormalization
group calculations and by a variational method.

Let us compare the results for $\delta_c$ at $J/t = 1/3$ in the honeycomb lattice
$(\delta_c^{hl})$ with those found in the square lattice $(\delta_c^{sl})$. Taking
$\delta_c^{hl}$ from the approach of~\cite{Gu13}, $\delta_c^{hl} \gtrsim 0.1$, and
$\delta_c^{sl}$ from the cumulant approach of~\cite{Vojta96}, $\delta_c^{sl} \simeq
0.045$, we have $\delta_c^{hl} / \delta_c^{sl} \simeq 2.2$. In the GMFA calculations we
get $\delta_c^{hl} = 0.069$ and $\delta_c^{sl} = 0.038$ (see~\cite{Vladimirov09}), i.e.,
the ratio $\delta_c^{hl} / \delta_c^{sl} \simeq 1.8$ nearly agrees with that given by the
approaches of~\cite{Gu13} and \cite{Vojta96}. That means, in the honeycomb lattice the
LRO is favored as compared with the square lattice, although the coordination number
$z^{hl} = 3$ is smaller than $z^{sl} = 4$. This behavior may be due to different
geometries of the hopping paths in the two lattices.

In Figure~\ref{fig5} the uniform static spin susceptibility $\chi$ is plotted as a
function of doping at various temperatures. The increase in $\chi$ upon doping is caused
by the decrease of AF SRO (cf. Figure~\ref{fig3}), i.e., of the spin stiffness against
the orientation along a homogeneous external magnetic field. At large doping, $\delta
\gtrsim 0.6$, $\chi$ decreases with increasing $\delta$ due to the decreasing number of
spins. Note that the position of the SRO-induced maximum of $\chi$ at $\delta_{max}(T)$
nearly agrees with the doping at which the further-distance correlation functions $C_n \,
(n=2,3,4)$ become vanishingly small (see Figure~\ref{fig3}). As compared with the
results for the $t-J$ model on the square lattice, the values of $\delta_{max}(T)$ are
much higher. That means, similar to the behavior of LRO, in the honeycomb lattice the SRO
is favored as compared with the square lattice.

Concerning the temperature dependence of $\chi$ at fixed doping, from Figure~\ref{fig5}
it can be seen that there appears a maximum at $T_{max}(\delta)$. This maximum can be
understood as a SRO effect in analogy to the explanation of the doping dependence of
$\chi$.

Figure~\ref{fig6} show the inverse correlation length $\xi^{-1}$. Considering the doping
dependence (Figure~\ref{fig6}(a) at $T = 0$, in the limit $\delta \rightarrow \delta_c
+$, AF LRO emerges which is connected with the closing of the AF gap, $\omega_+({\bf Q})
\rightarrow 0$, and by~(\ref{b20a}), with the divergence of $\xi$. At $T > 0$, there is
no LRO so that $\xi$ remains finite, where $\xi^{-1}$ almost linearly increases with
$\delta$. This corresponds to the weakening of AF correlations (see Figure~\ref{fig3}).
Let us consider the temperature dependence of $\xi$ (Figure~\ref{fig6}(b). At $\delta <
\delta_c$, $\xi$ diverges in the limit $T \rightarrow 0$, where at $\delta = 0$, $\xi$
exhibits the known exponential increase (see~\cite{Vladimirov17}).

The spectrum of spin excitations $\omega_{\pm}({\bf q})$ at $T = 0$ is shown in
Figure~\ref{fig7} along the symmetry directions $X(-1, 0) \rightarrow K (-2/3, 0)
\rightarrow \Gamma (0, 0) \rightarrow Y (0, 1) \rightarrow K (1/3, 1) \rightarrow \Gamma'
(1,1) \rightarrow M (1/2, 1/2) \rightarrow \Gamma$ of the BZ  shown in Figure~\ref{fig2}.
In the LRO phase, i.e., at $\delta < \delta_c$, the spin excitations are spin waves with
gapless branches depicted in Figure~\ref{fig7}(a). In the paramagnetic phase, spin waves
propagating in AF SRO can exist, if their wavelength is smaller than the correlation
length, i.e., if $q > q_c = 2 \pi \xi^{-1}$. For dopings slightly above $\delta_c$, where
the correlation length is large enough, this condition can be fulfilled. With increasing
doping, we may have $q < q_c$ so that the spin-wave picture breaks down, and "paramagnon"
excitations with the energies $\omega_{\pm}({\bf q})$ appear. Thus, our spin-excitation
spectra may reveal a smooth crossover from spin-wave to paramagnon behavior depending on
the wavenumber and doping. In the upper (optical) branch, at $\delta_c$ a gap is opening
at the $\Gamma$ point, i.e., at the AF wave vector ${\bf Q} = (0, 0)$ characterizing the
LRO phase in the two-sublattice model. As can be seen in Figure~\ref{fig7}, the
spin-excitation energies are decreasing with increasing doping. Interestingly, at the $K$
points the spin-excitation spectrum has a maximum.

\section{Conclusion}
\label{sec:6}

In this paper we have evaluated thermodynamic quantities as well as the electronic and
spin-excitation spectra in the strong-correlation limit within the $t-J$ model on the
honeycomb lattice. The dynamic spin susceptibility is calculated within a
spin-rotation-invariant generalized mean-field approach for arbitrary temperatures and
hole dopings. Our main focus was the analysis of the doping dependence of the
zero-temperature magnetization and of the uniform static spin susceptibility which we
have explained in terms of AF SRO. As compared with previous results for the $t-J$ model
on the square lattice, both the AF LRO and SRO are found to exist in a larger doping
region. We conclude that our investigation forms a good basis for forthcoming studies of
superconductivity in the honeycomb $t-J$ model.\\

\acknowledgments
The authors would like to thank V.~Yu.~Yushankhai and J.~Richter for valuable
discussions. The financial support by the Heisenberg-Landau program of JINR is
acknowledged. One of the authors (N. P.) thanks the Directorate of the MPIPKS for the
hospitality extended to him during his stay at the Institute.


\begin{thebibliography}{99}
\bibitem{Neto09} A. H. Castro Neto, F. Guinea,  N. M. R. Peres,
K. S. Novoselov, A. K. Geim, Rev. Mod. Phys. {\bf 81}, 109 (2009)
\bibitem{Kotov12}  V. N. Kotov, B. Uchoa, V. M. Pereira,
F. Guinea,  A. H. Castro Neto,  Rev. Mod. Phys.   {\bf 84}, 1067 (2012)
\bibitem{Sorella92} S. Sorella,  E. Tosatti, Europhys. Lett. {\bf 19}, 699 (1992)
\bibitem{Martelo97}  L. M. Martelo,  M. Dzierzawa, L. Siffert, D. Baeriswyl,
Z. Phys. B {\bf 103}, 335 (1997)
\bibitem{Furukawa01}   N. Furukawa, J. Phys. Soc. Jpn. {\bf 70}, 1483 (2001)
\bibitem{Peres04} N. M. R. Peres, M. A. N. Ara\'{u}jo, D. Bozi,
Phys. Rev. B  {\bf 70}, 195122 (2004)
\bibitem{Paiva05} T. Paiva, R. T. Scalettar, W. Zheng, R. R. P. Singh,  J.
Oitmaa,  Phys. Rev. B {\bf 72}, 085123 (2005)
\bibitem{Herbut06} I. F. Herbut,  Phys. Rev. Lett. {\bf 97}, 146401 (2006)
\bibitem{Herbut09}  I. F. Herbut, V. Juri\v{c}i\'{c},  O. Vafek,
 Phys. Rev. B {\bf 79}, 085116 (2009)
\bibitem{Raghu08} S. Raghu, Xiao-Liang Qi, C. Honerkamp,  Shou-Cheng Zhang,
Phys. Rev. Lett.  {\bf 100}, 156401 (2008)
\bibitem{Meng10} Z. Y. Meng, T. C. Lang, S. Wessel, F. F. Assaad,  A.
Muramatsu, Nature (London) {\bf 464}, 847 (2010)
\bibitem{Sorella12} S. Sorella, Y. Otsuka,  S. Yunoki, Sci. Rep. {\bf 2}, 992 (2012)
\bibitem{Aria15} S. Arya, P. V. Sriluckshmy, S. R. Hassan, A.-M. S. Tremblay,
Phys. Rev. B {\bf 92}, 045111 (2015)
\bibitem{Yang12} H.-Y. Yang, A. F. Albuquerque, S. Capponi, A. M. L\"{a}uchli,
K. P. Schmidt,  New J. Phys. {\bf 14}, 115027 (2012)
\bibitem{Lang13} T. C. Lang,. Z. Y. Meng, A. Muramatsu, S. Wessel,  F.F. Assaad,
Phys. Rev. Lett. {\bf 111}, 066401 (2013)
\bibitem{Assaad13} F. F. Assaad,  I. F. Herbut, Phys. Rev. X  {\bf 3}, 031010 (2013)
\bibitem{Toldin15} F. P. Toldin, M. Hohenadler,  F. F. Assaad,  I. F. Herbut,
Phys. Rev. B {\bf 91}, 165108 (2015)
\bibitem{Honerkamp08}  C. Honerkamp,  Phys. Rev. Lett. {\bf 100}, 146404 (2008)
\bibitem{Faye15}  J. P. L. Faye, P. Sahebsara,  D. S\`{e}n\`{e}chal, Phys. Rev. B {\bf 92},
085121 (2015)
\bibitem{Xu16}  X.-Y. Xu, S. Wessel,  Z.-Y. Meng, Phys. Rev. B {\bf 94}, 115105 (2016)
\bibitem{Luscher06} A. Luscher, A. Lauchli, W. Zheng,  O. P. Sushkov
Phys. Rev. B  {\bf 73}, 155118 (2006)
\bibitem{Gu13}  Z.-C. Gu,  H.-C. Jiang, D. N. Sheng, H. Yao, L. Balents,   X.-G. Wen,
Phys. Rev. B {\bf 88}, 155112 (2013)
\bibitem{Gu14} Z.-C. Gu, H.-C. Jiang, G. Baskaran, arXiv:1408.6820
\bibitem{Vladimirov17} A. A. Vladimirov, D. Ihle, N. M. Plakida,
Eur. Phys. J. B {\bf 90}, 48 (2017)
\bibitem{Vladimirov05} A. A. Vladimirov, D. Ihle, N. M. Plakida,
Theor. Mat. Phys. {\bf 145}, 1576 (2005)
\bibitem{Vladimirov09} A. A. Vladimirov, D. Ihle, N. M. Plakida,
Phys. Rev. B {\bf 80}, 104425 (2009)
\bibitem{Vladimirov11} A. A. Vladimirov, D. Ihle, N. M. Plakida,
Phys. Rev. B {\bf 83}, 024411 (2011)
\bibitem{Hubbard65} J. Hubbard,
Proc.~Roy. Soc. A (London) \textbf{285}, 542 (1965)
\bibitem{Plakida99} N. M. Plakida, V. S. Oudovenko, Phys. Rev. B  {\bf 59}, 11949 (1999)
\bibitem{Zubarev60} D. N. Zubarev,  Usp. Phys. Nauk \textbf{71}, 71
(1960) [translated in Sov. Phys. Usp. \textbf{3}, 320 (1960)]
\bibitem{Wallace47}  P. R. Wallace, Phys. Rev. {\bf 71}, 622 (1947)
\bibitem{Winterfeldt98} S. Winterfeldt, D. Ihle,
Phys. Rev. B {\bf 58}, 9402  (1998)
\bibitem{Low09} U. L\"{o}w,  Cond. Matter Phys. {\bf 12}, 497 (2009)
\bibitem{ST91} H. Shimahara, S. Takada, J. Phys. Soc. Jpn. {\bf 60},
2394 (1991)
\bibitem{WI97} S. Winterfeldt, D. Ihle, Phys. Rev. B {\bf 56}, 5535 (1997)
\bibitem{Junger09}  I. Juh\'{a}sz Junger, D. Ihle, J. Richter,
Phys. Rev. B  {\bf 80}, 064425 (2009)
\bibitem{Vojta96} M. Vojta, K. W. Becker, Phys. Rev. B {\bf 54}, 15483 (1996)
\end{thebibliography}
\end{document}